\documentstyle[epsf]{ptptex}

\newcommand{\sti} { \tilde{\tau}_1 }
\newcommand{\stii} { \tilde{\tau}_2 }
\newcommand{\cost}{\cos\theta_{\tau}}
\newcommand{\sint}{\sin\theta_{\tau}}
\newcommand{\stl}{ \tilde{\tau}_L}
\newcommand{\str}{ \tilde{\tau}_R}
\newcommand{\st}{\tilde{\tau}}
\newcommand{\msti}{m_{\tilde{\tau}_1}}
\newcommand{\mstii}{m_{\tilde{\tau}_2}}
\newcommand{\mstl}{m_{\tilde{\tau}_L}}
\newcommand{\mstr} {m_{\tilde{\tau}_R}}

\title{%
Precision Study of MSSM at future $e^+e^-$ linear colliders
}

\author{%
Keisuke {\sc Fujii}, Toshifumi {\sc Tsukamoto}, and Mihoko M. {\sc Nojiri}
\footnote{Talks given at Yukawa International Seminar(YIKS) '95
 on some very hot and humid day in August,
and also at Workshop on  {\it Physics and Experiments with Linear $e^+e^-$
Colliders} Appi, Iwate Japan Sep.8-12 1995.
E-mail address: nojirim@theory.kek.jp}
}
\inst{%
National Laboratory for High Energy Physics (KEK)\\
 Oho 1-1, Tsukuba, Ibaraki
305, Japan
}

\recdate{%
\today
}

\abst{%
The lighter scalar tau lepton
$\sti$  may be the lightest scalar lepton and
therefore would be found earlier in future collider experiments.
We point out the impact
of the measurement of the mass and the mixing angle of $\st$ to discriminate
the models of SUSY breaking. Furthermore, the measurement of the polarization
of $\tau$ lepton($P_{\tau}$) from the decaying $\sti$
helps to determine the Yukawa sector of minimal supersymmetric standard
model.
 We  present our   MC study of the production and the decay of $\sti$ lepton
at a future linear collider at $\sqrt{s}=500$ GeV.
The mass, mixing angle of $\sti$ and
 $P_{\tau}(\sti\rightarrow\tau\chi_1^0)$ would be measured precisely at
the future LC.
}

\begin{document}

\maketitle

\section{Introduction}
The Minimal Supersymmetric Standard Model (MSSM)\cite{rf:SUSY} is one of the
most
promising candidates of the models beyond the Standard Model (SM). It
predicts the existence of superpartners of SM particles below a few TeV to
remove quadratic divergence which appears in radiative corrections
of the SM Higgs sector;
thus the model is free from the so--called hierarchy problem of GUT
models. It should be noted that the gauge couplings unify very precisely at
high energy scale in MSSM,
SUSY SU(5) GUT predictions.

The supersymmetry is not an exact symmetry of the model,
instead it should be somehow broken to give the mass differences between
a particle and its superpartner. Various attempts have been made to
explain the existence of the soft SUSY breaking\cite{rf:SUGRA,rf:DNNS}.
Those different
models of SUSY breaking have different predictions for the
relation between the soft breaking mass parameters at some high scale $M_{SB}$;
 $m_i$ (scalar masses),
$M_i$(gaugino masses), $A_i$(trilinear couplings) and $B$( Higgsino
soft breaking mass parameter).
 Evolving the mass parameters by the  RGE of the model from
$M_{SB}$ to $M_{\rm weak}$,
one gets the prediction of the mass spectrum of superpartners
at the weak scale.

Therefore,
 the precise measurement
of masses and interactions of  superpartners will be
one of the most important physics targets once they are discovered.
This might enable us to discriminate
the models of even higher energy scale  responsible
for  the SUSY breaking if the experiment reaches certain sensitivity.
 Notice that to claim
a new particle as a superpartner also requires careful  investigations
of the interaction of the particle which should agree with
the expectations of supersymmetry.

 Proposed Linear Colliders
at $\sqrt{s}=500$ GeV are  expected to have high
luminosity---${\cal L}=30 fb^{-1} $/year\cite{rf:JLC1,rf:LC}.
The background from $W$ boson production can be suppressed drastically
thanks to the highly polarized electron beam; Current technology
already archived $P_{e^-}=80\%$ at SLC, and $P_{e^-} =95\%$ is
proposed at future LC's.
Under this clean environment, precision study of the mass and
interaction become possible.
Studies of accelerator technology  for the future LC's
are on going in several institutes such as SLAC, KEK,  DESY and
CERN.\cite{rf:LC}

The potential impact of a LC to the supersymmetric models
have been pointed out by several groups already\cite{rf:TSUKA,rf:FLC}.
 For example, the predictions of the Minimal Supergravity(MSUGRA) model for
$M_1/M_2$ and $m_{\tilde{e}}/$ $m_{\tilde{\mu}}$ have been shown
 to be proven up to ${\cal O}(1\%\sim 10\%)$.
 Ino-lepton-slepton coupling also can be measured to check the
prediction of supersymmetry.
Those analyses have been done for
$\tilde{e},\tilde{\mu}$ and $\tilde{\chi}^+$ pair production modes.

In the following, I will talk about our MC study of
the production and decay of $\tilde{\tau}$ at a
future LC\cite{rf:NO,rf:WIP}.
The decay of $\tilde{\tau}$ involves a $\tau$ lepton, which decays
further in the detector. It makes analysis rather complicated,
and therefor MC study of the process have not been done previously. However,
the physics coming out from the study turns out to be fruitful, due to
the unique nature of $\st$ interaction through Planck scale to
the weak scale.

In Sec. 2.1, we briefly describe the reduction of $m_{\st_{L,R}}$
by the GUT scale  Yukawa interaction in
MSUGRA-GUT model, which has been pointed out recently by
Barbieri and Hall\cite{rf:BH}.
$\st$ would
be found earlier than the other SUSY particles in the model as
the $\st$ mass is expected to be much lighter
than other sleptons.
It is also stressed measurement of $m_{\st_{L,R}}$ provides
clear cuts to distinguish the MSUGRA-GUT from  other models.

Blow the GUT scale, the interaction of $\tau$ lepton is still
different to the other sleptons as it has a non-negligible Yukawa coupling
$Y_{\tau}\propto m_{\tau}/\cos\beta$; Here $\tan\beta$ is the ratio
of vacuum expectation values of the two neutral Higgs boson in MSSM.
The Yukawa coupling is enhanced
linearly $\propto\tan\beta$ for large value of $\tan\beta$.
A consequence of the
large Yukawa coupling is existence of left-right mixing of $\st$;
The lighter mass eigenstate of $\tilde{\tau}$  would be lighter than
the other sleptons even if mass parameter of $\st$ is equal to that
of $\tilde{e}$ and $\tilde{\mu}$. The feasibility of  determination of
the mass and mixing angle at a future LC is checked by
MC simulation in Sec. 2.2.

The same $Y_{\tau}$ appears as a non-negligible $\tau\tilde{\tau}
\tilde{H}^0_1$ coupling, where $\tilde{H}^0_1$ is a neutral higgsino.
The ratio of the couplings involving higgsino component
and gaugino component of neutralino $\chi^0$,
where the neutralino is a mixture of higgsinos and gauginos,
can be determined through the measurement of  the polarization of
$\tau$ lepton($P_{\tau}$) from $\st$ decay into a neutralino and $\tau$.
The strong sensitivity of $P_{\tau}$ to
$\tan\beta$ helps  to determine $\tan\beta$,
by combining the information from the other modes.
 The performance of LC experiment on the determination
 of $P_{\tau}$ will be found in Sec. 2.3.
Sec. 3 is devoted for conclusion and discussions.

\section{Study of Scalar Tau Lepton at LC}
\subsection{ Mass of Scalar Tau and Models of Supersymmetry breaking}
$\tilde{\tau}_{L(R)}$ is the superpartner of $\tau_{L(R)}$,
the third generation lepton. This makes $\tilde{\tau}$
an unique object in the context of the SUGRA-GUT model\cite{rf:SUGRA}.
In the supergravity model, the SUSY breaking in the hidden sector
gives the soft breaking mass through gravitational interaction
at Planck scale  $M_{pl}$. The resulting scalar
mass is universal at $M_{pl}$, leading to approximate
universality of $m_{l_{L,(R)}}$
if their interaction is equal from $M_{pl}$ to $M_{\rm weak}$.
However, in simple grand unified models such as SO(10) or SU(5),
the $\tau$ superfield is in the same multiplet with
the top quark superfield above the GUT scale $M_{GUT}$.
Thus from $M_{pl}$
to  $M_{GUT}$, the $\tau$
supermultiplet obeys the same Yukawa interaction as that of top quark.
The large top Yukawa interaction is anticipated by
the top mass measurement by CDF or D0\cite{rf:TOP},
 and this reduces the masses of
$\tilde{\tau}_R$ ( or $\tilde{\tau}_{L(R)}$)
 at $M_{GUT}$ compared to its value at $M_{pl}$ for SU(5)
(or SO(10)) GUT model. This is pointe out in  Ref.\citen{rf:BH} and
they claimed that $m_{\tilde{\tau}}$
can be as light as a half of $m_{\tilde{e}}$.
$m_{\tilde{\tau}}$ may even be the second
lightest SUSY particle in this model.

I should stress that there exists a model
 which predicts totally different mass
spectrum. Dine-Nelson-Nir-Shirman\cite{rf:DNNS} recently
constructed relatively
simple models which
break SUSY at an intermediate scale
 [ $\sim10^{6\sim 7}$ GeV] dynamically(DNNS model).
The breaking is then transformed to our sector by $U(1)$ gauge interaction,
which is called a messenger sector. The scale where the gauge interaction
breaks($M_{m}$)  is $O(10^4)$ GeV.
Due to the nature of the gauge interaction,
the resulting scalar masses of sleptons are common for ($l_L, \nu_l$) and
$l_R$ at $M_{m}$ respectively. Unlike SUGRA-GUT model,
they remain roughly equal at $M_{weak}$, as
$M_{m}$ is considerably close to
$M_{weak}$ and there is no strong Yukawa interaction involved
between the scales.  Therefore, determination of $m_{\tilde{\tau}_{L,R}}$
would give us a good handle to distinguish the scale of
SUSY breaking below or above the GUT scale.

\subsection{ Determination of $\st$ mass matrix at LC}
To determine $m_{\tilde{\tau}_{L,R}}$, one has to
know  $\tilde{\tau}$ interaction. This is because
neither $\tilde{\tau}_L$ nor $\tilde{\tau}_R$ is a mass eigenstate, but
they generally mix to make the mass eigenstates $\tilde{\tau}_{1(2)}$;
The mass matrix is expressed as
\begin{subequations}\label{eq:1}
\begin{equation}
{\cal M}^2_{\st}=\left(\begin{array}{cc}m_{LL}^2 & m_{LR}^2\\
m_{LR}^2& m_{RR}^2\end{array}\right)
=\left( \begin{array}{cc}
m_L^2 + m_{\tau}^2 + 0.27 D &   -m_{\tau}(A_{\tau} + \mu \tan\beta)\\
 -m_{\tau}(A_\tau + \mu \tan\beta)&  m_R^2 +m_{\tau}^2 + 0.23D
 \end{array}\right),\\
\label{eq:1a}
\end{equation}
and the mass  eigenstates  are  expressed as
\begin{equation}
\left(\begin{array}{c} \sti\\\stii\end{array}\right)
=\left(\begin{array}{cc}\cost &\sint\\
 -\sint&\cost\end{array}\right)
\left(\begin{array}{c} \stl\\ \str\end{array}\right).
\end{equation}
\end{subequations}
Here $\mu$ is Higgsino mass parameter, $\tan\beta\equiv\langle H_1^0\rangle
/\langle H_{2}^0\rangle$ is the ratio of vacuum expectation values, and
$A_{\tau}$ is the coefficient of the soft breaking term proportional to
 $\tau_R$-$\tau_L$-$H_1$, and $D$ corresponds to $D$-term.

The mixing makes the lighter mass eigenvalue $\msti$ lighter than
diagonal mass terms, thus even in the model with the common
soft breaking scaler mass, $m_{\sti}$ may be
lighter than $m_{\tilde{e}}$\cite{rf:DN}.
At the same time,  one has to know
$\theta_{\tau}$ together with $\msti$ and $\mstii$
to determine $\mstl$ and $\mstr$.
Notice that it is interesting to observe
the non-zero $\theta_{\tau}$(mod $\pi$)
as this proves the existence of the off-diagonal element
of the $\sti$ mass matrix;
this  depends on the term proportional to $\mu\cdot\tan\beta$ which is
required from supersymmetry, while $A_\tau$ is the
coefficient of the trilinear soft breaking term.
Both terms are strongly motivated by the supersymmetric theory.

If the electron beam is polarized, the mixing angle $\theta_{\tau}$
will be determined from the measurement of
the production cross section
 $e^+e^-\rightarrow\sti^+\sti^-$\cite{rf:NO}.
This can be easily explained by taking
the limit where $m_Z\ll\sqrt{s}$ and $P_e=1$.  In the limit, the
production of $\st$ solely proceed through $U(1)$ gauge interaction
that carries hypercharge. The hypercharge is $-1/2$($-1$)  for
$\tilde{\tau}_{L(R)}$, thus $\sigma(\tilde{\tau}_R)$
$\sim 4 \sigma(\tilde{\tau}_L)$. The cross section also depends
of $\msti$, however this would be extracted from the energy
 distribution of $\st$ decay products, as we will see later.

To show the feasibility of the measurement of $\msti$ and $\theta_{\tau}$
at a future $e^+e^-$ collider, we did MC simulation for the JLC1 detector
\cite{rf:JLC1}.
We took  $\sqrt{s}=500$ GeV and $P_e=95\%$, and analysed the process where
$\sti$ decays into $\chi_1^0\tau$ exclusively; here
$\chi_1^0$ is the lightest neutralino and  we assumed $\chi_1^0$ to be
the lightest SUSY particle and stable, and  we denote it by $\chi$ hereafter .

Due to the simple  2 body kinematics, the energy distribution of
$\tau$ leptons is flat between
$E_{min}$ to $E_{max}$ , which contains the information about $m_{\chi}$
and $\msti$. Actually for the process $e^+e^-\rightarrow \tilde{e}^+
\tilde{e}^-$ and $\tilde{e}\rightarrow \chi e$, the energy distribution of
the electrons  was used to determine
$m_{\tilde{e}}$ and $m_{\chi}$\cite{rf:TSUKA}.
However  the $\tau$ lepton decays further into $\pi,\rho,a_1, e$ and $\mu$
etc..
The decay distribution depends not only on the $E_{max(min)}$
but also  on the decay modes of
$\tau$ lepton.
we reconstructed $\rho$ and $a_1$ whenever it is possible.\footnote{
For Monte Carlo simulation, we used TAUORA ver2.4\cite{rf:TAUORA}. See next
subsection to our cuts to identify $\rho$ and $a_1$.}

We require both of the $\tau$ decays hadronically as  signal
to avoid relatively large  background
from $eeZ^0$ and $e\nu W$.
We also include
backgrounds  from $W^+W^-$, $Z^0Z^0$, $e^+e^-W^+W^-$  and $\nu\nu Z^0$
productions.  Cuts like
$E_{vis}>10$ GeV and $\theta_{accop}>30^{\circ}$
are applied to reduce the backgrounds too. The resulting signal
of $\tau$ production
is characterized as 2 jets of low hadron multiplicity with missing $P_T$.
Those selected MC samples are then used to `measure'
 $m_{\chi}$ and $m_{\tilde{\tau}_1}$ by fitting
the energy distribution of the MC sample.

In figure 1, we show the results of the mass fit for the
sample identified by a tau leptons decaying into
$\rho$.  Here we generated 10,000 $\st$ pairs with
mass $\msti=150$ GeV which decayed into $\chi\tau$ with $m_{\chi}=100$ GeV.
We also included
backgrounds  consistent with  $\int L=100fb^{-1}$.
About 1700 events of $\rho$ are obtained for the signal after the
cut, while 93 events remained as the backgrounds.
Contours of constant $\chi^2=-1/2 \log L$ are shown in Fig. 1.
We show the result in
the  $m_{\chi}$ and
$m_{\sti}$ plane fixing the other parameters
\footnote{The results are obtained  by normalizing
 the total number of the events of the fitting curve of the signal
and background by the  event number
obtained by MC.  $P_{\tau}=1$  both for MC and the theoretical distribution.
The determination of $P_{\tau}$ is discussed in the next subsection}.
$\msti$ is determined with the error of 3.5 GeV.
The error of the cross section at the best fit point is 2.5\%.

The errors corresponding to 5000 events are shown
in fig. 2 schematically in
$m_{\sti}$ and $\sin\theta_{\tau}$
plane(scaled statistically),
where the contours of constant
$\sigma_{\sti}$(=$50fb$ dotted line, $50\pm 1.25 fb$ solid lines)
are shown simultaneously. $\delta\theta_{\tau}=\pm 4.5^{\circ}$
is read from the figure.

We showed the measurement of the $\sti$ production and decay
 can determine two of the three parameters of $\st$ mass matrix.
Discovery of $\stii$ would specify the remaining degree of freedom.



\subsection{ The Yukawa sector of MSSM and $P_{\tau}$}

The study of $\tilde{\tau}$ may play important role in exploring the
Yukawa sector of MSSM\cite{rf:NO}.  Let's  consider the decay of
$\sti\rightarrow \tau\chi_1^0$ again.  The $\chi_1^0$ is the mixture of
gauginos ($\tilde{B},\tilde{W}$) and Higgsinos ($\tilde{H}_{1(2)}$ ).
The interaction involving gaugino component
($\tilde{B}(\tilde{W})$-$\tau$-$\st$ coupling) is
proportional to gauge couplings and the interaction involving
Higgsino component ($\tilde{H}_1$-$\tau$-$\st$ coupling)
 is proportional to $\tau$ Yukawa coupling
$Y_{\tau}\sim m_{\tau}/\cos\beta$.
The latter may not be
too small compared to the former when $\tan\beta$ is large or
$\chi_1^0$ has the large higgsino component(See Fig. 3).

Those two interactions are different not only in the couplings, but
also in the chirality of the (s)fermion. The (super-) gauge
interaction is chirality conserving, while the (super-) Yukawa
interaction flips it. ( In Fig.3, the arrows next to the $\st$ and $\tau$
lines show the direction of chirality.) Thus the polarization of $\tau$
lepton ($P_{\tau}$) from $\sti$ decays
depends on the ratio of the chirality flipping and concerning
interactions.

$P_{\tau}(\sti\rightarrow\tau\chi^0_1)$ depends
on $\tan\beta$  strongly compared to other
quantities. To demonstrate this, we show various quantities in
Fig.4 a)-d) fixing $m_{\chi^0_1}=100$ GeV and varying $M_1$($\tilde{B}$
mass parameter) and $\tan\beta$.
 Fig.4 a)-c) show little dependence on $\tan\beta$.
Especially  the pair production of $\tilde{e}_R$ can be
used as the mode to determine $M_1$ as in Ref.\citen{rf:TSUKA}.
On the other hand,  $P(\tilde{\tau}_R\rightarrow\tau\chi^0_1)$
depends on $\tan\beta$ sensitively if $M_1$ is sufficiently larger
than $m_{\chi^0_1}$ or $\tan\beta$ is large.
This is because the chirality flipping higgsino interaction
becomes comparable to the chirality conserving gaugino interaction
either if the lightest neutralino is dominantly Higgsino or
if $Y_{\tau}$ is large
\footnote{ The neutralino sector is parametrized by $M_{1(2)}$,
$\mu$, $\tan\beta$. When $\mu\ll(\gg )M_1$, $\chi^0_1$
is dominantly higgsino(gaugino)
and $m_\chi\sim\mu (M_1)$. In Fig.4, $\chi^0_1$
becomes dominantly higgsino as $M_1$ becomes larger.
}.
In such a situation, one can determine
$\tan\beta$ by using the value of $M_1$ obtained by the
other production processes.\footnote{One can also determine
$\tan\beta$ from forward-backward asymmetry of
chargino production cross section. It is sensitive
to $\tan\beta$ when $ M_2\sim \mu$ \cite{rf:FLEP,rf:FLC}}

The measurement of $P_{\tau}$ would be carried out through the
energy distribution of decay products from the polarized $\tau$
lepton. The $\tau$ lepton decays into $A\nu_{\tau}$ where
$A=e, \mu,  \pi, \rho, a_1$... 
For the each decay channel,
the momentum distribution of the hadronic decay products
($\pi^{\pm}$, $\rho^{\pm}\rightarrow\pi^{\pm}\pi^0$...)
differs significantly
depending on $P_{\tau}$. If the $\tau$ lepton is relativistic,
 $P_{\tau}$ can be determined from the energy distribution of
the decay products\cite{rf:HAGI}.

Being more specific, let us consider the decay of a polarized  $\tau$ lepton
into $\rho$. The $\rho$ from right(left) handed  $\tau$ lepton
is longitudinally(transversally) polarized. The $\rho$ meson then decays
into $\pi^{\pm}\pi^0 \rightarrow \pi^{\pm}2\gamma$. The energy fraction
$ E_{\pi^{\pm}} /E_{\rho}$, where $E_{\rho}$ is the total energy of jets which
the $\pi^{\pm}$ belongs to, depends on $\rho$ polarization in a very
simple form  in
the collinear limit( $E_{\tau}\gg m_{\tau}$);
\begin{equation}\label{eq:2}
\frac{d\Gamma(\rho_T\rightarrow 2\pi)}{dz}
\sim 2z(1-z)-\frac{2m_{\pi}^2}{m_{\rho}^2},\ \
\frac{d\Gamma(\rho_L\rightarrow 2\pi)}{dz}
\sim (2z-1)^2,
\end{equation}
where $z=E_{\pi}/E_{\rho}$.

Notice that decay of the  $\tau$ lepton into the heavier meson
$a_1$ may be misidentified as decay into $\rho$ if
the energy and momentum resolution of the detector
is poor. The decay $\rho\rightarrow\pi^+ 2\gamma$ will be identified as
a jet of a $\pi^+$ and one or two photon candidates. On the other
hand, the decay $a_1^{\pm}\rightarrow \pi^{\pm}\pi^0\pi^0\rightarrow
\pi^{\pm}4 \gamma$ is also occasionally  misidentified as
$\pi^{\pm} 2\gamma$ or $\pi^{\pm}\gamma$  which
contaminates  $\rho$ signals. We applied
the cuts $m_j<0.95$ GeV for events with one photon candidate
and $m_j<0.95$ GeV and $m_{2\gamma}<0.25$ GeV for events
with two photon candidates
 to reduce the contamination from  $a_1$ decay;
after the cut the contamination is less than a few \% for
$m_{\sti}=150$ GeV $m_{\chi}=100$ GeV.
However, the purity of the sample
 crucially depends on the assumed performance
of the JLC 1 detector\cite{rf:JLC1}. If this is not achieved, one may
have to look up the decay mode into $\tau^{\pm}\rightarrow
\pi^{\pm}\nu$ or $a_1^{\pm}\rightarrow \pi^{\pm}\pi^{\pm}\pi^{\mp}$.
The branching ratio into those modes are  small
compared to the one into $\rho$.

Fig.5. shows the $z$ distribution of MC events and the fit
for the same parameter with Fig 1.
The best fit value of $P_{\tau}$ is $0.95 (-0.92)$ for $P_{\tau}=1(-1)$
respectively and the estimated error is $\pm 0.07$
\footnote{For the analysis we took the event where $0.08<z<0.92$ to avoid
 detector effects. We used the events $E_j>20 $ GeV as the events
below the cut does not have the sensitivity to  $P_{\tau}$}

\section{Conclusion}
I presented in this talk our study of the production and decay of the lighter
scaler tau lepton $\sti$ at a future LC. The study of $\st$ is
important because $\st$ may be lighter than the
other sleptons, thus would be found earlier.
The light $\sti$ is well motivated in MSUGRA-GUT model,
and it is not excluded in other models, if there is large
$\tilde{\tau}_L$-$\tilde{\tau}_R$ mixing.

We discussed that the  mass matrix of $\st$ provides
a clue to distinguish   SUGRA-GUT model and DNNS model, and
the  polarization of $\tau$ lepton $P_{\tau}$
from  decaying $\sti$  is sensitive to
the value of $\tan\beta$ through its dependence to the $\tau$
Yukawa coupling; $\tan\beta$ is one of the important parameters
to determine the Higgs sector of the MSSM.

The feasibility of the study of those parameters  at the LC have been  checked
by MC. The error of $\msti$ and $\sigma(\sti\sti)$
(which in turn used
to determine the mass matrix of $\sti$) and $P_{\tau}$ are
3.5 GeV, 2.5 \% and 0.07 respectively for
a representative parameters we have chosen.
 We have not  included several potentially
important background such as $\gamma\gamma\rightarrow \tau\tau$ and
production and decay of heavier superpartners. However, we believe final
results will not be too much different to the ones we have presented here.

 In near future, LEPII and LHC are scheduled to operate at
$\sqrt{s}=180$ GeV  and $\sqrt{s}=14$ TeV respectively.
However, their ability to determine the soft breaking mass parameters
is rather poor. For LEPII, integrated luminosity is ${\cal O}(100 pb^{-1})$,
while the production cross sections  are typically ${\cal O}$(10pb)
for chargino and  $0.3$ pb for $\tilde{\mu}$ with
$m_{\tilde{\mu}}= 60$ GeV. The production cross section
of a slepton is too small to go beyond discovery physics.
 Feng and Strasslar showed
that precise study of chargino interactions is possible\cite{rf:FLEP}, however,
one has to still fight over the enormous background coming
from $W^+W^-$ production. SUSY study at LHC(Large Hadron Collider)
suffers from the high QCD background although strongly interacting
superpartners will be copiously produced at LHC.
Expected signals of SUSY particle production
are also very complicated, as decay patterns of squarks and
gluino change drastically depending on the  mass spectrum of SUSY particles.

Implications of MSUGRA model at LEPII and LHC  have been discussed
and studied in quite a few papers.
Those are mostly about the
reduction of the number of free parameters of the model (which
tighten the phenomenological constraint), ``theoretical upper bound'' of
sparticle masses, and (therefore) when and how they would be discovered.
However, it is becoming recognized that we can go beyond that if
a next generation LC is actually built. Namely, the experiment at the LC
will make it possible to measure the parameters of MSSM once a superpartner is
discovered, and this enablesy us to check the predictions of the models of
SUSY breaking. I presented in this talk that discovery of $\st$ at the LC
provides us a clear cut to understand  the origin of SUSY breaking, and
I hope I have convinced the audiences that the LC is necessary to achieve that.

\section*{Acknowledgements}
We would like to thank Y. Okada and B. K. Bullock for careful reading
of manuscripts.

\end{document}